\documentclass[10pt]{iopart}

\usepackage{graphicx}
\usepackage{titlesec}
\usepackage{xcolor}

\definecolor{red2}{RGB}{209,35,42}
\definecolor{blue2}{RGB}{11,125,180}
\titlelabel{\thetitle.  \quad \quad \quad}
\titleformat{\section}   
{\color{red2}\normalfont\large\bfseries} 
{\color{red2}\thesection.}{0.25em}{} 

\begin{document}
\title{\textsf{Relative  Stability of Bernal and Rhombohedral Stackings in Trilayer Graphene under Distortions}}

\author{R Guerrero-Avil\'es$^{1}$ , M Pelc$^{2,3}$, F R Geisenhof$^{3}$, T Weitz$^{4}$, and A Ayuela$^{1}$}

\address{$^{1}$ Material Physics Center CFM-MPC, Donostia International Physics Center (DIPC), Paseo Manuel Lardizabal 4-5, 20018 Donostia-San Sebasti\'an, Spain}
\address{$^{2}$ Institute of Physics, Nicolaus Copernicus University in Toru\'n, Grudziadzka 5, 87-100 Toru\'n, Poland}
\address{$^{3}$ Physics of Nanosystems, Department of Physics, Ludwig-Maximilians-Universit{\"a}t M{\"u}nchen, Amalienstrasse 54, 80799 Munich, Germany}
\address{$^{4}$ I. Physikalisches Institut, Friedrich-Hund-Platz 1, 37077 G\"ottingen, Germany
}

\ead{
     \mailto{\color{blue2} swxayfea@sw.ehu.es}} 

\vspace{10pt}
\begin{indented}
\item[]{Keywords: few-layer graphene, rhombohedral stacking, Bernal stacking, deformation, stability\/}
\end{indented}

\begin{abstract}
\noindent Stackings in graphene have a pivotal role in properties to be discussed in the future, as seen in the recently found superconductivity of twisted bilayer graphene. Beyond bilayer graphene, the stacking order of multilayer graphene can be rhombohedral, which shows  flat bands near the Fermi level that are associated with interesting phenomena, such as tunable conducting surface states expected to exhibit spontaneous quantum Hall effect,  surface superconductivity, and even topological order.However, the difficulty in exploring rhombohedral graphenes is that in experiments, the alternative, hexagonal stacking is the most commonly found geometry and  has been considered the most stable configuration for many years.Here we reexamine this stability issue in line with current ongoing studies in various laboratories. We conducted a detailed investigation of the relative stability of trilayer graphene stackings and showed how delicate this subject is. These few-layer graphenes appear to have two basic stackings with similar energies. The rhombohedral and Bernal stackings are selected using not only compressions but anisotropic in-plane distortions. Furthermore, switching between stable stackings is more clearly induced by deformations such as shear and breaking of the symmetries between graphene sublattices, which can be accessed during selective synthesis approaches. We seek a guide on how to better control – by preserving and changing - the stackings in multilayer graphene samples.
\end{abstract}
\maketitle
\ioptwocol  
\section{\color{red2}Introduction}
\indent Graphene remains a key two-dimensional material to still successfully reveal intriguing properties when stacked into just a few layers. As an example, bilayer graphene in twisted form has been found to exhibit superconductivity properties \cite{Cao2018}.
Large graphene samples are obtained experimentally by mechanical exfoliation of graphite, so they are supposed to favor the Bernal stacking, which for graphite has been considered more stable 
than the rhombohedral stacking  \cite{Tomanek1988}. However, this stability is far from being fully settled. Firstly, recent calculations showed that graphite is 
energetically more stable in the rhombohedral stacking, so it is suggested that crystal growing conditions under high temperatures and pressures in a geological scale can stand behind graphite in the Bernal stacking \cite{xiao2011density}.
Secondly, experimental and synthesizing techniques in 2D materials have been so improved over the years that graphene is being investigated nowadays with a defined number of layers \cite{Sugawara2018,Park2016}.
Research works that focused on the same exfoliated sample of trilayer graphene (TLG) simultaneously found regions with the Bernal and rhombohedral stacking \cite{Lui2011,Xu2012,Cong2011,Warner2012,geisenhof2019anisotropic}. Rhombohedral graphenes are today being synthesized with a revival in a research race to show unique properties mostly due to their flat bands \cite{shi2020electronic}. The stability between the Bernal and rhombohedral stacking in few layer graphenes therefore stands as a crucial question to be discussed.

Techniques are currently being developed on how to obtain, preserve or change stacking in trilayer graphene (TLG). The phase change between rhombohedral (ABC) and Bernal (ABA) stacking can be induced by external driving forces. 
The ABC stacking was experimentally found when graphene layers were grown on SiC substrates, and it was stable up to a temperature of around 1,200 C$^\circ$ 
when it reverts to the ABA stacking \cite{Ferrari2006,Graf2007}.
The transition from ABA to ABC stacking can be caused by a scanning tunneling microscope (STM) tip on highly ordered pyrolytic graphite (HOPG) samples due to the small barriers involved ($\sim1.0$ mev/atom), as recently reported \cite{Xu2012,yankowitz2014electric}. 
The transition between ABC and  ABA stacking was also achieved using triazine decoration, which causes a large energy difference to induce the stacking transition \cite{Zhang2013}.
The stacking in few-layer graphenes was stabilized mechanically by strain, using electric fields and even doping \cite{yang2019stacking,gao2020large,li2020global}.
In order to fabricate high-quality encapsulated ABC trilayer - hBN devices used for quantum transport investigations, it has been shown that often the ABC trilayer converts to ABA during van-der-Waals transfer, even though ABA regions had been carefully removed prior to stamping \cite{Chen2019}. The rhombohedral to Bernal transition has been theoretically studied by modeling the continuous sliding deformation of a side layer and by parameterizing results for two layers, calculations that were performed within density functional theory (DFT) using local density approximation (LDA) \cite{aoki2007dependence,nery2020}. Although the van-der-Waals energies per atom are small, the total bonding energies are sizeable when the sample areas are considered,  total energies that are large but needed to keep a certain phase locked in a metastable stacking after being synthesized.
These findings suggest that not only the relative stability between these TLG stackings should be studied, but that their dependence on the lattice deformations deserves further study.

In this work, we address the stability between rhombohedral and Bernal stacking  in trilayer graphene by calculations performed within density functional theory (DFT) using explicitly van der Waals functionals. We study different deformations that induced the energetic exchange between these stackings by looking at the relative stability.
We first show that as the TLG ground state, the rhombohedral stacking is slightly more stable than the Bernal one.
Then we perform several deformations in the quest to induce the transition from rhombohedral stacking to the Bernal one.
In our simulations we find that the Bernal stacking is energetically more stable than the rhombohedral one under some distortions based on the planar expansion of the lattice parameters and their perpendicular compression. Furthermore, we show that under distortions such as shear and the atomic shifts in graphene sublattices, the structural anisotropy helps to stabilize  the ABA stackings in certain directions. 
These mechanisms point out that the stacking transition is achieved by breaking the sublattices symmetries of layer-layer interactions involved in each stacking.
These findings indicate that different practical realizations are possible, such as depositing samples in sandwiches, substrates, and molecule decoration, where stacking transitions may take place.

\section{\color{red2}TLG stability and electronic structures}
\begin{figure}[hbpt]
	\centering
	\includegraphics[width=8cm]{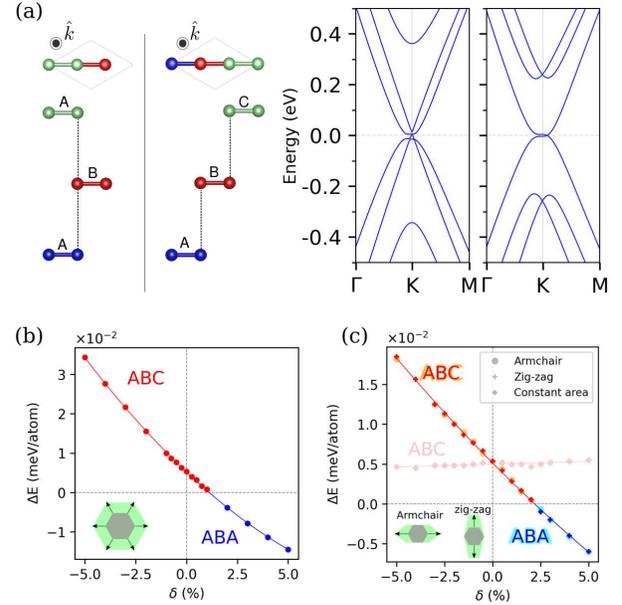} 
	\caption{(a) (Left) Side views of the units cells and (right) band structures for the  Bernal and rhombohedral stacking of trilayer graphene (TLG). 
		(b) Energy differences  between Bernal and rhombohedral TLG versus in-plane homogeneous strains $\delta$ with compression ($\delta<$0) and stretching ($\delta>$0) in lattice parameters. 	
		(c) Energy differences for uniaxial deformations along zigzag and armchair directions. Blue-cyan and red-orange colors refer to the strain values when the Bernal and rhombohedral  stackings are more stable, respectively. Energy differences are also given with strains assuming a constant area per atom, shown in light red. The insets in panels (b) and (c) indicate the homogeneous and uniaxial deformations in-plane, respectively. The TLG rhombohedral stacking is more stable without surface strain, and a transition to Bernal stacking is shown under expansions with $\delta \geq 1\%$. 
	}
	\label{fig1}
\end{figure}

Figure \ref{fig1}(a) shows the relaxed unit cells of trilayer graphene with  the Bernal stacking on the left and the rhombohedral stacking on the right.
Each stacking has a relaxed intralayer  atomic distance of $a_{c-c}=1.43$\AA, and the interlayer distance of $d=3.55$\AA\ in agreement with experiments \cite{kittel1996introduction}. The relative stability between Bernal and rhombohedral stackings is next analyzed by looking at the difference between their total energies.
After geometry relaxations, the energy differences per atom are calculated as $\Delta E = (E_{B}-E_{r})/N$, where $E_B$ and $E_r$ are the total energies of Bernal and rhombohedral stacking respectively, and  $N=6$ is being the number of atoms in the unit cells.
For pristine TLG our former results show a total energy difference of 0.079 meV/atom between both stackings. 
We have performed a detailed analysis of converging parameters to the accuracy required here, as shown in supplementary data. It seems theoretically established that the Bernal stacking is more stable than rhombohedral stacking in the literature, but this statement is based on inconsistent results \cite{xiao2011density}. In fact, we find that in TLG the rhombohedral stacking is slightly more favorable than the Bernal one.
 The energy differences although small are in agreement with enthalpy experiments 
 between hexagonal and rhombohedral graphene
 \cite{li2020global,boehm1964enthalpy}.  Kinetic conditions such as temperatures and pressures seem to be the key in obtaining the Bernal stacking. All atoms have to shift the entire layer in order to pass from ABA to ABC stacking, and the other way around. We propose  that these processes have to be considered in order to compare with the room temperature energy taking into account the total contact area. This means that there is a minimum area in the flakes   as to obtain a stacking stable at room temperature. Depending on the value of stacking energy times area, there is a blocking temperature above which the separate layers can slide. In other words there is also a minimum area at a given temperature so that the flake stacking remain stable to be observed in experiments. An implication is that stackings can be different depending on the synthesis methods  using exfoliation and CVD growth.   
We compare the electronic band structures of the Bernal and rhombohedral stackings along the $\Gamma$KM path near the Fermi energy, as shown in Figure \ref{fig1}(a). At the K point the Bernal stacking bands are displaying superposed a massive and massless electronic behavior, and the rhombohedral electronic structure shows massive electrons only; these two types of dispersion bands for each stacking are in agreement with Ref. \cite{Menezes2014}.
The Bernal stacking can be seen as a superposition of single layer graphene bands and Bernal bilayer graphene bands; however, the graphene-like bands are slightly shifted up indicating the source of some intrinsic electric field between the layers to be added to external fields \cite{lui2011observation}.
For the rhombohedral stacking there is a parabolic behavior with a flat region around the K point, source of many interesting properties, such as  tunable conducting surface states \cite{shi2020electronic} expected to exhibit spontaneous quantum Hall effect \cite{zhang2011spontaneous},  surface superconductivity \cite{kopnin2013high}, and even topological order \cite{slizovskiy2019films}. 
The massless band in the Bernal stacking now has acquired mass in the rhombohedral geometry. The conduction and valence bands are degenerated and slightly displaced from the K point in the KM direction. At lower energies, the rhombohedral stacking presents split bands what justifies its stability found in the calculations.
A simple image can allow us to explain this stability. The Bernal stacking has directly connected three atoms between layers so that there are states at zero energy. However, the rhombohedral case has two pairs of atoms between layers so the corresponding electronic states are split from the zero energy, a fact that could be somehow linked to its higher stability.

\section{\color{red2} In-plane deformations}
We now study different in-plane deformations of the lattice vectors. 
Each deformation is characterized by a stretch  factor $\lambda$ that operates in the lattice vectors on different ways.
First, we consider a homogeneous (or isotropic) deformation of lattice vectors described by $\vec{a}_{1,2}  = \lambda(a_{x}\hat{i}+a_{y}\hat{j})$.
In this sense, having $\lambda$=1 (or strain $\delta=0\%$) shows the previous relaxed lattice vectors, while a value of $\lambda=1.05$ indicates an expansion strain of 5$\%$ ($\lambda=0.95$ is a compression of $5\%$).
This causes a proportional deformation of the lattice vectors keeping the interlayer distance constant. 
Figure \ref{fig1}(b) shows the total energy differences per atom between Bernal and rhombohedral stacking for  the in-plane deformations explained above.  Positive values of the energy differences (colored in red) indicate that the  rhombohedral stacking is more stable than the Bernal one. 
On the contrary, negative values for the energy differences (colored in blue) refer to the case of the Bernal stacking being more stable.
We show that although under homogeneous deformation the Bernal stacking becomes more favorable on expansion, the energy difference decreases for positive strains till it changes sign for $\sim 1 \%$. At this point the Bernal stacking becomes more favourable.

Next, we consider anisotropic deformations with uniaxial strain without and with constant area, described by the lattice vectors $\vec{a}_{1,2}  = \lambda a_{x}\hat{i}+a_{y}\hat{j}$  ($\vec{a}_{1,2}  =  a_{x}\hat{x}+\lambda a_{y}\hat{y}$) and $\vec{a}_{1,2}  = \lambda a_{x}\hat{i}+a_{y} / (\lambda)\hat{j}$ along $\hat{x}$ ($\hat{y}$) direction, respectively.
The energy difference induced by the anisotropic deformation considering a constant area is slightly decreasing and remains almost constant in the whole range of considered positive or negative strains in the order of 5  \%, as shown in Figure \ref{fig1} (c). Furthermore, this figure  includes the energy differences with the anisotropic uniaxial deformations, and we find that the Bernal stacking becomes more stable during stretching.
The values of energy in the strain range $\pm 5 \%$ show smaller response compared with those of homogeneous deformation, 2 versus 5 tenths of meV.
The rhombohedral stacking with anisotropic strains is more stable for strain values till $\delta\sim2.2\%$, to be compared with
the Bernal stacking with isotropic strains that becomes more stable under small strains $\delta\sim1.1\%$ .\

\begin{figure}[hbtp]
\centering
\includegraphics[width=8cm]{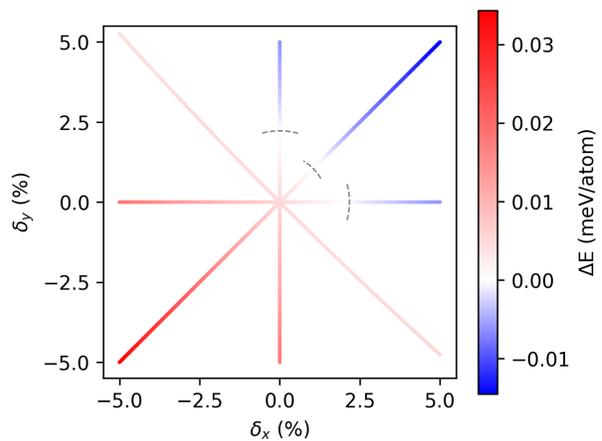}
\caption{ Summary of energy differences with respect to in-plane deformations $\delta_x$ and $\delta_y$. Note the anisotropy of the rhombohedral-Bernal stability versus in-plane deformations.}
\label{fig2}
\end{figure}

Figure \ref{fig2} summarizes the trends found between the Bernal and rhombohedral energy differences versus the in-plane deformations in TLG. There is an asymmetry in strains that induces that the most stable rhombohedral stacking switches into the Bernal stacking. The strain values when this transition is taking place are within the experimental range of reversible strains attained for samples in substrates, already found for  a single layer of graphene \cite{kim2009large}. In fact, strain values of beyond 1\% could be artificially applied to graphene heterostructures \cite{Wang2019} and strain up to 0.3\% has been seen in graphene encapsulated in hBN \cite{Vincent2018}. 
Thus detailed care must then be taken in experiments when depositing exfoliated graphenes on substrates, and when using patterned contacts \cite{geisenhof2019anisotropic,bouhafs2021synthesis}, 
so that strain is induced by top contacts while edge contacts can help in reducing strain \cite{Sanctis2018}.

\section{\color{red2}Out of plane deformations}
\begin{figure}[hbtp]
	\centering
	\includegraphics[width=8cm]{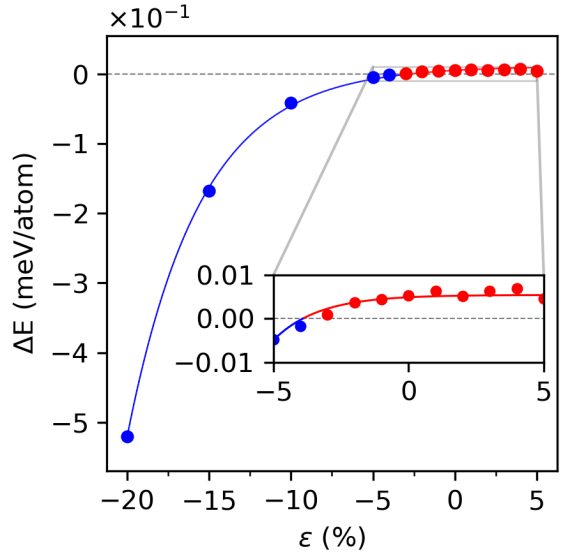}
	\caption{ Out of plane deformations: total energy differences versus  out-of-plane strain $\epsilon$ related to the interlayer distance. 
	} 
	\label{fig3}
\end{figure}

We further study the relative stability order induced by the strain $\epsilon$ in the interlayer distance by stretching and compression between layers, as shown in Figure. \ref{fig3}.
Although the energies of both stackings are almost degenerated at zero strain,
$\Delta E$ under stretching shows that the rhombohedral continues being more stable even for an interlayer distance up to $d=4.26$\AA\ ($\epsilon\leq 20\%$).
For a compression at $d=3.11$\AA \ ($\epsilon=-2.5\%$), the $\Delta E$ values show that  the Bernal stacking becomes more favorable, in agreement with another current calculation using empirically fitted vdW Grimme functional \cite{nery2021ab}.
Small compressions induced by processes during transfer and protective layers will be enough to end into the ABA stacking.  While we are not aware of direct measurements of compressions during h-BN encapsulation or subsequent cleaning by squeezing out residues \cite{Purdie2018}, one can anticipate that compression of a trilayer can take place during these processing steps.

The next question is to determine how under a perpendicular pressure
also the atomic in-plane layers are modified.  
It is noteworthy that under constant volume a perpendicular pressure implies that the layers are expanding on the plane. For instance assuming a constant volume, a value of $\epsilon=-3.5\%$ 
corresponds to a homogenous expansion of $\delta=1.88\%$. The critical $\delta$ value reinforces that the transition to Bernal stacking comes coupled to the in-plane deformations, as shown  in Figure \ref{fig1}(b) and (c) above.  
These results indicate that the TLG stackings have to be analyzed in the sample once they are covered and isolated in the operating device whenever is possible to be certain that the rhombohedral stacking has not changed into Bernal.   
\section{\color{red2} Shear strain deformations}
To assess the role of symmetry breaking we analyze the shear strain deformations.
We consider the shear deformations along the perpendicular layer direction using the azimuthal $\phi$ and polar $\theta$ angles, as shown in the inset of Figure \ref{fig4} and the movie included in supplementary data.
The question is what $\theta$ values must be chosen before entering into  different stackings from the rhombohedral and Bernal ones, i.e. how close the structures with shear can be seen as deformations of the original ABA and ABC stackings. The shear structures remain close to the pristine ABC and ABA stackings for the values with $\theta < 15^\circ$, as shown in supplementary data, where shears beyond this critical angle show layers that can hardly be seen anymore as small deformations of the Bernal and rhombohedral stackings.

Figure \ref{fig4} shows energy differences between Bernal and rhombohedral stackings for the shears with $\theta=10^\circ,5^\circ$ changing $\phi$ angles.
The shear for $\theta=10^\circ$ also provokes that the interlayer distance \emph{d}  decreases from the value in pristine lattices so that $\epsilon=-1.5\%$; however, this $\epsilon$ value is still well below the critical distance to get the ABA stacking with perpendicular pressure. So that when obtaining Bernal stackings is not just correlated with the distance but better with the shear angle.
Note that for $\theta \neq  0^\circ$ the shear is displacing the layers in the in-plane with respect to the perfect stacking as cones, shown in the upper panel, so that the overlap of the carbon p$_z$ orbitals between layers change related not only to distances but also to angles. In fact  the $\epsilon$ value for $\theta=5^\circ$  ($\epsilon\sim 0.3 \%$) just decreases slightly the interlayer distance, being larger than the critical distance to obtain ABA stacking under pressure. 
\begin{figure}[hbtp]
	\centering
	\includegraphics[width=8cm]{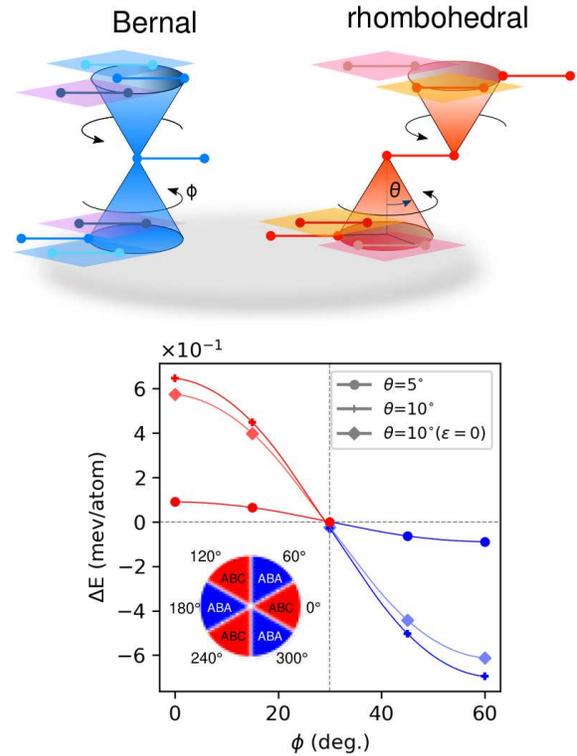}
	\caption{Stability energy difference between Bernal and rhombohedral stackings versus shear deformation, which is indicated by angles $\theta$ and $\phi$ shown in the upper panel. The inset shows the shear $\phi$ angles for which the ABA and ABC stackings are more stable.  
	}
	\label{fig4}
\end{figure}
On the one hand, the energies of Bernal stacking exhibit a $60^\circ$ periodicity in $\phi$, while increasing $\theta$ makes the energy profile just wider. 
That means that these deformations destabilize Bernal stacking minimally up to $\phi=30^\circ$.
On the other hand, the energies for rhombohedral stacking does not show the same periodicity along $\phi$ angle, because their periodicity (120$^{\circ}$ in $\phi$) is twice than one found for  Bernal stacking.
Thus we find a region where the rhombohedral stacking is more destabilized than the Bernal one.

In Figure \ref{fig4} we show the energy difference per atom between the two stackings under shear deformations. 
The curve for $\theta=5^\circ$ shows the $\Delta E$ in the range of $\sim\pm0.1 meV/atom$  decreasing as $\phi$ values increase.
The case of $\theta=10^{\circ}$ where the $\Delta E$ are in the range of $\sim\pm6$ tenths of meV/atom much above the hundreths of meV/atom found for previous studied deformations; in fact they become an order of magnitude larger.
For values of $\phi$ less than $30^\circ$ the rhombohedral stacking still remains as more favorable, and for the values of $\phi>30^\circ$, we find that the Bernal stacking is more stable than the rhombohedral one.  Taking into account the symmetry, we find that the Bernal stacking becomes favorable for three fold directions of $\phi=60^\circ$ as shown in the inset included in Figure \ref{fig4}. Interestingly, the ABA stackings for $\theta=5^{\circ}$ are observed even for very small interlayer distance, values with very small $\epsilon$. Even for a test case with $\epsilon=0$, i.e with no change in interlayer distance, the energies remain in the order of tenths of meV.
In general, considering that those $\Delta$E  values are larger than previous considered deformation cases, the shear deformation of the stackings promote the Bernal stacking in certain directions,  so that the area of stacking regions can even be reduced by a factor ten to be stable using shear deformations.
These findings are in line with previous experimental results in which shear is induced by contacts \cite{geisenhof2019anisotropic}, by other encapsulating layered graphenes within BN  \footnote{Along zigzag and armchair direction the ABC and ABA stackings were favored, respectively} \cite{yang2019stacking}, and by explicitly putting shear to the samples  \cite{jiang2018manipulation}.  
\section{\color{red2} Perpendicular sublattice  distortions}
\label{disp}
Up to now all the investigated deformations affected both graphene sublattices within the same layer in the same way. We have then shown that shear introduced mostly anisotropy in plane. For the sake of completeness we now asses the role of deformations out of plane by looking at the anisotropy induced by the sublattices sites. 
We study the displacement up and down for each of the sublattices in both stackings.
We characterize how much the same perpendicular shift affects the relative stability of each stacking.
Here, we apply a perpendicular displacement to A or B sublattice type nodes, shifts that are breaking the sublattice layer symmetry. We shift an atom of a particular sublattice in the direction perpendicular to the layer plane by less than the layer compression to get ABA stacking, and compare how much the total energy of each stacking is changed.
\begin{figure}[hbtp]
	\centering
	\includegraphics[width=6cm]{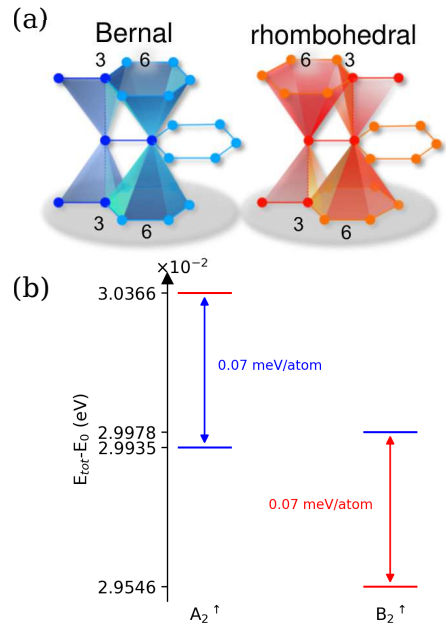}
	\caption{ 
		(a) Models for the Bernal and rhombohedral stacking including the number of interlayer next nearest neighbors of the middle layer atoms, to be discussed in text. (b) Comparison of relative energies with respect the ground state of the perpendicular displacements of the A$_{2}^{\uparrow}$ and B$_{2}^{\uparrow}$ graphene sublattices in the middle layer. The subscript denotes layer number, and the superscript arrow indicates the corresponding up and down shift of the sublattice atom.} 
	\label{fig5}
\end{figure}

First we note that in each stacking, atoms of both side layers have the same number of interlayer neighbors. The energy response for atom shifts in side layers is very small, it does not exceed 0.01 meV. Thus we focus on the middle layer analyzing the coordination numbers of its atoms. Figure \ref{fig5} (a) shows the Bernal and rhombohedral stacking schemes indicating the nearest neighbours in next layers. The vertical lines are linking the middle layer atoms to the first neighbors belonging to the side layers and shaded prims with numbers mark the second interlayer neighbors. In both stackings the A and B type nodes have in total 2 first neighbors and 18 second neighbors belonging to the side layers. However, the number of neighbors for each sublattice calculated separately is different. We can expect different responses when breaking the symmetry by shifting A or B atoms. 
In the case of Bernal stacking the numbers of interlayer neighbors for a sublattice site are the same with top and bottom layer. Thus the up or down shifts will be symmetric.
In the rhombohedral stacking the numbers of neighbors for each sublattice site are ``antisymmetric'' i.e. shifting the A atom up is equivalent to shift the B type atom down 
and vice versa. For the rhombohedral geometry the number of cases to be discussed is reduced to two: (i) $A_2^\uparrow$ equivalent to $B_2^\downarrow$ and (ii) $A_2^\downarrow$ 
equivalent to $B_2^\uparrow$, where subscripts refer to the layer number, and arrows indicate the perpendicular displacement.

Figure \ref{fig5}(b) shows the comparison of energy response difference defined as $E_{shift} - E_0$, where for each stacking in the two types of displacements $E_0$ is the ground state energy, and $E_{shift}$ is the total energy when the atoms are shifted. We shift the atoms by 1\% of interlayer distance, i.e. 0.0335\AA. In all the considered cases the total energy in each stacking increases by $\approx$ 0.5 meV. The energy response due to each sublattice shift remains nearly the same for the ABA stacking, and for the ABC stacking it changes more depending on the shift direction. For ABC stacking, moving the A atom up - equivalent to shift the B atom down - causes different energy response in the two directions. In the case of rhombohedral stacking, the energy increase is higher by 0.07 meV. As a result the Bernal stacking becomes more stable. The situation is just opposite when moving the A atom down (equivalent to shift the B atom up) - in this case the Bernal stacking is more destabilized, and the rhombohedral geometry remains favourable. The perpendicular sublattice shift studied here is a lattice deformation that causes an energy difference between stackings about 0.07 meV. This finding points out that breaking the sublattice symmetry is another way to stabilize the Bernal and rhombohedral stackings.

These energy differences due to breaking of sublattice symmetry can be explained by having the TLG stackings interpreted again in terms of trimers and dimers between layers. The total energy can be written in two parts as a sum of eigenvalues and a sum of interatomic potentials with other atoms. Looking at the total contribution of interatomic potentials the second neighbours atoms in the nearest layers are held responsible for these energy differences. This energy contribution adds to the difference in the sum of eigenvalues, already commented, in which the levels for the case of Bernal stacking split from the zero energy when the atoms are shifted up or down. These calculations raise further intringuing experiments regarding how the sublattice symmetry of stackings could be considerably broken by substrates and adatoms, or even better by having molecules and nanoparticles conforming these graphene heterostructures \cite{zhao2014growth,huang2020large}.

\section{\color{red2} Conclusions}
The aim of the present paper was to determine when the graphene stacking changes from rhombohedral to Bernal due to small lattice deformations. 
Under compression  the rhombohedral stacking suffers a phase transformation to Bernal with strains $\epsilon < 2.5 \%$. For even smaller compressions, shear deformations and shifts breaking graphene sublattice symmetries induce an anisotropy that stabilizes the Bernal stacking. 
These findings provide insights into the role of substrate-associated strains when graphene layers are integrated into devices, so that the stacking order and consequently their ultimate electronic properties are modified. This work would be of interest in relevant technological areas such as patterning contacts, and encapsulating graphene flakes between other materials. Our results are then asking for further experiments looking at the role of directional shears along and the adsorption of adatoms and molecules on graphene stackings.  

\ack{This research was funded by the Spanish Ministry of Science and Innovation (grants no. PID2019-105488GB-I00), the Gobierno Vasco UPV/EHU (project no. IT-1246-19), and the European Commission NRG-STORAGE project (project no. GA 870114). 
This research was conducted in the scope of the Transnational Common Laboratory (LTC) “Aquitaine-Euskadi Network in Green Concrete and Cement-based Materials."}

\section*{\color{red2}Methods}\label{compudetail} 
We performed our study within density functional theory (DFT) using the ab-$initio$ simulation package VASP \cite{Kresse1993,Kresse1996,Kresse1996a}. This method based on plane waves is applied using a well converged kinetic energy cut-off of 700 Ry. 
In the energy range that we are interested,  the mesh grid on the reciprocal space is key to get consistent results (see supplementary data), a  288$\times$288$\times$1 centered at the $\Gamma$ point is used. 
For the dispersive interactions between layers, we perform our calculations using the Van der Waals functional  vdW-DF2 \cite{Klimes2009}. 
The electronic convergence was performed with $10^{-7}$ eV. 
We relax the cell shape and keep the volume constant, and we fix the ions in the $xy$ plane while allowing them to move in the $z$ direction. 
Tests performed with other functionals, as well as detailed convergence test in k-points, and electronic and ionic relaxations are included in supplementary data. 
Note that a fine k-mesh is required to be applied with the number of sampling k points in the x and y directions being a multiple of 3 to explicitly include the points K, K\'\ and a region around them, regions in the reciprocal space where the low energy physics of multilayer graphene occurs.

\section*{\color{red2}Supplementary data}
The following files are available free of charge.
\begin{enumerate}
    \item[$\bullet$] Convergence tests to properly simulate few-layer graphene. Tests using Several vdW Functionals. Farthest shear deformations in Bernal and rhombohedral stackings.
    \item[$\bullet$] Movie to visually follow the shear deformations of trilayer graphene stackings.  
\end{enumerate}
This material is available free of charge via the Internet at https://iopscience.iop.org/.




\section*{References}
\bibliographystyle{iopart-num}
\providecommand{\newblock}{}

\end{document}